\documentclass[superscriptaddress,aps,preprint,longbibliography]{revtex4-1}
\usepackage{graphicx} 
\usepackage{amsmath}
\usepackage{amssymb}
\usepackage[utf8]{inputenc}
\usepackage{booktabs}
\usepackage{enumitem}
\usepackage{xcolor}
\graphicspath{{./pictures/}}
\begin{document}
\title{Direct test for critical slowing down before Dansgaard-Oeschger events via the volcanic climate response}
\author{Johannes Lohmann}
\email{johannes.lohmann@nbi.ku.dk}
\affiliation{Physics of Ice, Climate and Earth, Niels Bohr Institute, University of Copenhagen, Denmark}

\begin{abstract}
It is tested whether past abrupt climate changes support the validity of statistical early-warning signals (EWS) as predictor of future climate tipping points. EWS are expected increases in amplitude and correlation of fluctuations driven by noise. This is a symptom of critical slowing down (CSD), where a system's recovery from an external perturbation becomes slower as a tipping point (represented by a bifurcation) is approached. 
EWS are a simple, indirect measure of CSD, but subject to assumptions on the noise process and measurement stationarity that are hard to verify. 
In this work the existence of CSD before the Dansgaard-Oeschger (DO) events of the last glacial period is directly tested by inferring the climate's recovery from large volcanic eruptions. By averaging over hundreds of eruptions, a well-defined, stationary perturbation is constructed and the average climate response is measured by eight ice core proxies. As the abrupt DO warming transitions are approached, the climate response to eruptions remains the same, indicating no CSD. For the abrupt DO cooling transitions, however, some key proxies show evidence of larger climate response and slower recovery as the transitions are approached. 
By comparison, almost all proxies show statistical EWS before cooling and warming transitions, but with only weak confidence for the warming transitions. 
There is thus qualitative agreement of CSD and EWS, in that the evidence for bifurcation precursors is larger for the cooling transitions. 
However, the discrepancy that many proxies show EWS but no direct CSD (and vice versa) highlights that statistical EWS in individual observables need to be interpreted with care. 
\end{abstract}

\maketitle

\section{Introduction}

Earth's history has seen abrupt changes in stability of different climate sub-systems that resulted in in catastrophic regime shifts \cite{LEN08,DAK08,ROU23}. These are also referred to as climate tipping points (TP) and may be manifestations of dynamical bifurcations. Do to the risk of climate TP in the future as a result of global warming it is of great interest to predict from data whether we are approaching such changes. 
This may be possible because the climate dynamics prior to such a dynamical bifurcation is expected to display critical slowing down (CSD) \cite{HAK80,NOR81,WIS84,CRO88,KLE03,SCH09}, 
i.e., a characteristic decrease towards zero of the system's recovery rate after perturbations as the regime shift is approached. Since climate sub-systems are furthermore exposed to random fluctuations in their boundary conditions or some other external forcing, CSD gives rise to an increase in the amplitude and correlation of natural variability. These changes in the fluctuations of the unperturbed system can be used as statistical early warning signals (EWS), which is currently being exploited to warn about potential ongoing or impending climate stability shifts in the context of global warming \cite{BOE21,BOE21a,BOU22,MIC22,DIT23,BOC23}. 

The paleoclimate record is a key testbed for the validity and practical skill of EWS. The Dansgaard-Oeschger (DO) events of the last glacial period \cite{DAN93} are some of the best studied recurring past abrupt climate changes, and are a prime candidate for this due to their good preservation in climate proxies of high-resolution ice core records.
EWS prior to DO events have been analyzed before in $\delta^{18}$O temperature proxy records from Greenland ice core records, with mixed results ranging from no EWS \cite{DIT10}, to some indication of EWS when averaging over all DO events \cite{CIM13}, to significant EWS in a substantial number of the events when focusing on multi-decadal frequency bands \cite{RYP16,BOE18}. A recent study finds on the contrary a decrease in variability at decadal frequencies in the new EGRIP ice core \cite{BRA25}. 

\begin{figure}
\includegraphics[width=0.99\textwidth]{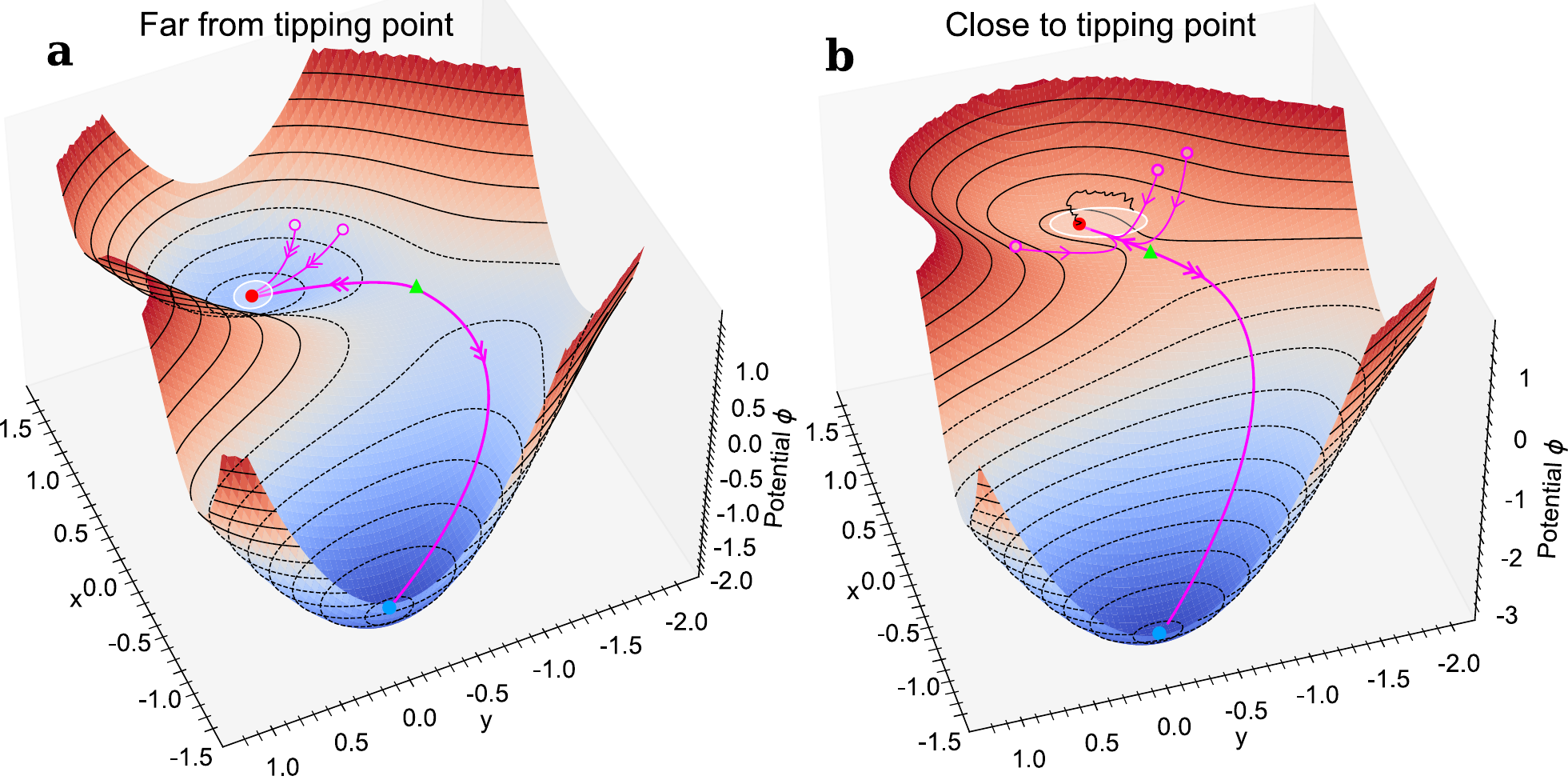}
\caption{\label{fig:ews_schematic} 
Critical slowing down and early-warning signals for a tipping point (TP) caused by a saddle-node bifurcation, illustrated for dynamics in a two-dimensional (quasi-)potential. 
There are two stable fixed points at local minima of the potential (red and blue dots), as well as one saddle point (green triangle). When far from the TP ({\bf a}), both minima are relatively deep. After applying fast perturbations to the system (open circles), it relaxes quickly back to the fixed point. Noise-driven fluctuations are small and relatively isotropic, indicated by the small white ball around the red fixed point. 
When close to the TP ({\bf b}), one minimum is very shallow and the potential around it very flat. The most flat direction is along the slow (center) manifold, which is the line connecting the saddle and the fixed point. 
Now, relaxation trajectories after a perturbation first quickly approach the manifold and then evolve very slowly back towards the fixed point. This is CSD. Noise-driven fluctuations become much larger, which gives rise to EWS, and evolve in a preferred direction indicated by the white ellipse. If the system is non-gradient, i.e., does not evolve solely in directions defined by the gradient of a potential, 
large fluctuations do not have to be directed towards the center manifold, as illustrated by the short noisy trajectory in black. 
}
\end{figure}

The present study addresses two shortcomings of statistical EWS in this context. First, EWS are only indirect indicators of CSD, and may be masked due to non-equilibrium effects and changes in strength and correlation of the process generating dynamical and observational noise, as well as multiplicative noise \cite{PRO23,MOR24}. The climate is thermodynamically out-of-equilibrium, and thus a non-gradient dynamical systems with non-fixed point dynamics. This means there are oscillatory modes that generally change in an unknown way as the bifurcation is approached, thus interfering with EWS intended to measure fluctuations around fixed points \cite{TAN18,LOH24,LOH25}. 
Further, under the action of random perturbations from the environment the non-gradient system becomes a non-reversible stochastic process. This can lead to differences CSD and EWS. CSD is a slowing of the relaxation back to an underlying attractor of the deterministic system after discrete perturbations, and arises because the system forms a slow (center) manifold as the bifurcation is approached (Fig.~\ref{fig:ews_schematic}). 
Statistical EWS measure the stochastic motion away from the attractor. While in a gradient (equilibrium) system this motion is aligned with the slow manifold, in a non-gradient system large fluctuations may be directed along different directions. Hence, it is not guaranteed that observables strongly projecting on the center manifold (thereby capturing CSD) also display significant EWS. This is especially true if the random environmental fluctuations do not excite the relevant degrees of freedom \cite{BOE13}, and if data contains observational noise that is not part of the system's natural, dynamic fluctuations. 

Instead of relying on statistical EWS, it is thus more robust to directly infer CSD from repeated controlled perturbations. Such an approach has been taken in experiments on biological systems \cite{DAI12,VER12,DAI13}, but has not been attemped for climate data. In the present paper it is argued that controlled climate perturbations can be analyzed
if one considers averages over enough instances of specific, fast-acting natural perturbations. Volcanic eruptions provide relatively abrupt perturbations to the atmospheric radiative balance, after which the climate system responds and relaxes back to its prior state. The climate forcing from one large eruption to the next is vastly different due to different eruption latitudes, seasonality, and other factors. 
But averages over dozens or hundreds of large eruptions can be considered a well-defined, stationary forcing, which on average induces a cooling of the climate and subsequent changes in atmopshere, ocean and biosphere. 
The average climate response to the eruption average should be stationary if the stability of the climate system remains unchanged. Conversely, a potential change in the stability as a regime shift is approached may be measured by changes in the average response to eruptions. 
To perform this averaging across several hundred perturbations, a record of volcanic eruptions obtained from ice core records of the last glacial period is used here \cite{LIN22}.
Considering average responses filters out variability that is not synchronized to the (randomly occurring) volcanic eruptions, such as observational noise in climate proxies. The approach also does not suffer from changes in the dynamical noise process or its multiplicativeness.

The second shortcoming to be addressed is that the critical dynamics may not project well on individual climate records. Hence, statistical EWS on an a priori chosen observable may be masked, even though CSD in the system as a whole may be present. The variability changes leading to EWS may only be visible in certain degrees of freedom that are directly related to the physical mechanism of the bifurcation \cite{LOH25}. 
As a result, the analysis of individual climate records in order to find bifurcation precursors for past or ongoing stability shifts can be inconclusive or misleading. Indeed, past work on the abrupt climate changes during the last glacial period has focused on the analysis of a singe ice core proxy record ($\delta^{18}$O), with the exception of the inclusion of a dust record in \cite{BOE18}. 
In the present work, a suite of eight high-resolution proxy records from a single ice core (the NGRIP ice core \cite{NGR04}) is considered in order to cover different modes of climate variability in the assessment of CSD and statistical EWS. 
Further, from this multivariate proxy set a scalar observable is extracted that should best carry the critical dynamics and EWS. This is done via the dimension reduction method of diffusion maps \cite{COI05}, which approximates the eigenfunctions of the adjoint of the operator that governs the probability density of the stochastic-dynamic climate system. The first non-trivial eigenfunction defines an observable that is expected to display the critical dynamics \cite{LUC24,LOH25b}. 

A last improvement with respect to prior work is made by estimating to what degree statistical EWS in the ice core records can be considered significant given that the proxy recording process is non-stationary due to changing snow accumulation rate and ice flow. 

The remainder of the paper is structured as follows. In Sec.~\ref{sec:methods} the climatic and volcanic ice core data sets are detailed, and the preprocessing of the time series is discussed.
Sec.~\ref{sec:volc_signal} discusses the average volcanic signal in the different proxies. 
Sec.~\ref{sec:volc_csd} then gives the analysis of the CSD signal in the multivariate proxy record in response to volcanic eruptions. In Sec.~\ref{sec:ews} statistical EWS in the different proxies are assessed, and Sec.~\ref{sec:ews_significance} presents the analysis of significance of these signals in the light of non-stationarity in the ice core proxy archive. Discussion and conclusions are given in Sec.~\ref{sec:discussion}.

\section{Methods and Materials}
\label{sec:methods}

\subsection{High-resolution ice core data}

All proxy time series are from the NGRIP ice core \cite{NGR04}, and comprise a $\delta^{18}$O oxygen isotope record \cite{GKI14}, a layer thickness record as proxy for accumulation rate changes \cite{RAS23}, mineral dust concentrations \cite{RAS23}, as well as concentrations of five chemical species ($NO_4$, $Ca$, $Na$, $NH_4$ and $SO_4$) as soluble impurities measured by continuous-flow analysis \cite{ERH22, LIN22}. The investigated time period is 11.7-60 ka. This covers 22 DO cycles, starting with the onset of interstadial GI-17.2 (see \cite{RAS14} for nomenclature) and ending at the termination of GS-1 (the Younger Dryas), which is also the onset of the Holocene. 
Greenland stadials (GS) refer to the cold phases of DO cycles, whereas Greenland interstadials (GI) refer to the relatively mild phases. 
In Fig.~\ref{fig:timeseries} the time series of $\delta^{18}$O and $Ca$ are shown, as well as the division of the record into DO cycles comprising stadials and interstadials. In the following, the records are explained in more detail, in particular how they are brought from an uneven sampling in terms of ice core depth at different resolutions onto an evenly spaced time grid on the GICC05 age scale. 

The dust and chemical species records were measured in 1 mm depth resolution \cite{ERH22,RAS23}, and after transferring the measurement depths to the Greenland Ice Core Chronology 2005 (GICC05) age scale the time series were downsampled by linear spline interpolation to an equidistant 0.1 year grid. The measurement units for dust are counts (from a laser microparticle detector) per milliliter and parts per billion for the chemical species. In all plots we show minus the logarithm thereof. For $\delta^{18}$O, due to the lower measurement resolution a 1-year equidistant grid is used as in \cite{LOH24c}, which was obtained by linear interpolation of the midpoint depths of the measurement intervals onto GICC05 time-depth scale, and subsequent oversampling of this unequally spaced time series to a 1-year equidistant grid using nearest-neighbor interpolation. 

The layer thickness record is obtained from the annual layer-counting of the NGRIP core \cite{RAS23}, which has been performed until 60.2 ka BP and includes certain and uncertain layers. For certain layers, a depth layer increment corresponds to a one year time increment. In uncertain layers (10.1\% of all layers) subsequent depths are defined as a half-year time increment \cite{AND06}. To obtain the record, the depth-age pairs of the GICC05 chronology are converted to thickness-age pairs by taking the increment of subsequent depths. Then, to homogenize the record containing full and half years, it is linearly interpolated to a 0.1 year grid. This does not represent the actual accumulation rate, since the flow-induced layer thinning is not accounted for. But on the time scales of several millennia or less the modulation by ice flow is quite small (see Sec.~\ref{sec:ews_significance}). Thus, for the analysis in this paper of the multi-annual volcanic anomaly and short-term changes in the proxy variability, the layer thickness $\lambda$ can safely be treated as direct measure of annual accumulation rate. 

\subsection{Data set of volcanic eruptions}
\label{sec:methods_eruptions}

The data set of volcanic eruptions is published in \cite{LIN22}, and covers the period 11.7-60 ka with $N=780$ eruptions identified in the NGRIP core. The eruptions were identified based on their sulfur deposition in the ice core above a certain threshold, which corresponds to about half the deposition of the Tambora eruption in 1815 CE. On average, they are thus expected to significantly perturb the climate system, and they indeed induce a clear anomaly in the ice core climate proxies \cite{LOH24c}. This motivates why they are suitable to study the post-eruptive relaxation of the climate back to its steady state. 
The corresponding eruption ages on GICC05 are as in \cite{LOH24c}, which corresponds to a slight adjustment compared to \cite{LIN22} (very minor for NGRIP), and a shift of the ages back in time by 1.5 years relative to the time of maximum sulfate deposition, to account for the fact that the maximum sulfate peak in the ice core is delayed with respect to the eruption age \citep{BUR19}. 

In this work only the 'stable' parts of the stadials and interstadials are considered, by excluding the actual abrupt DO warming and cooling transitions that last on average about 63 and 70 years, respectively \cite{LOH19}. 
As a result, 34 eruptions from the original data set are discarded. Estimates for the timing and duration of DO transitions are taken from previous studies. The onsets of the DO warming transitions, i.e., the ends of stadials have been estimated precisely from a stacked (using several ice cores) Greenland $\delta^{18}$O record in \cite{LOH22}. The other required anchor points, i.e., the beginnings of stadials, as well as the beginnings and ends of the interstadial periods are defined by the piecewise-linear fitting technique from \cite{LOH19} on the same stacked Greenland $\delta^{18}$O record. These estimates have also been used in \cite{LOH22} for the onsets of the DO cooling transitions. Since here the data from all DO cycles is aggregated, and furthermore larger intervals before the abrupt transitions are averaged, the exact timings do not affect the results much. 

\subsection{Time series preprocessing}
\label{sec:methods_timeseries}

In the study of the volcanic climate response (Sec.~\ref{sec:volc_csd}), for each proxy record short anomaly time series segments around the volcanic eruptions are taken and then averaged across many eruptions to filter out noise and non-volcanic climate variability. 
For each volcanic eruption, a 100-year slice centered around the eruption year is taken and linearly detrended. Thereafter, the mean of the 50 years preceding the eruption is removed to make sure the baseline climatological anomaly prior to the eruption is zero. This yields time series segments for each volcanic eruption that are then lumped together in different groups according to their age relative to the DO transitions (see Sec.~\ref{sec:volc_csd}), and finally averaged to obtain a mean anomaly. 

For the study of statistical EWS (Sec.~\ref{sec:ews}), i.e., the climate variability not only around volcanic eruptions, all proxy time series are brought on the same 1-year equidistant grid as $\delta^{18}$O, which has been done by averaging. The record is then cut into the GI and GS periods comprising the DO cycles, and each GS or GI is further divided into $n$ segments of equal length. Here, $n=5$ is chosen as a trade-off between good statistics and time resolution. These segments represent climate conditions that are progressively closer to the purported bifurcation of the abrupt DO transitions. The results are also tested on high-pass filtered data, constructed by convoluting the data with a Gaussian kernel with width of 50-years and subsequently subtracting the convoluted signal from the original data. 

In Sec.~\ref{sec:ews} a scalar observable is constructed from the eight proxy records using the diffusion map algorithm \cite{COI05}. This algorithm is suited for finding the degree of freedom with slowest relaxation dynamics \cite{COI08,LOH25b}, i.e., a function of the eight proxy variables as an observable that can capture CSD. The correct degree of freedom may only be expressed when relatively close to a bifurcation. Thus, for each GS and GI only the third of the data that is closest to the abrupt DO warmings and coolings is used. Ideally one would restrict to segments even closer to the transition, but there is a trade-off with data size to achieve a good approximation. The algorithm does not require time ordering of data points, and thus all time series segments of the different DO cycles are concatenated to yield one data set for GI and GS with $N=5882$ and $N=9282$ data points, respectively. 
For the diffusion map algorithm, all pairwise distances in the eight-dimensional space of the data points are fed into a Gaussian kernel of bandwidth $\epsilon$, and from this a Markov matrix is constructed that defines a random walk on a weighted graph connecting the data points. $\epsilon = 5$ has been chosen here. An eigendecomposition of the matrix is performed, and the first non-trivial eigenfunction (eigenvalue closest to 1) defines the desired observable (see Sec.~\ref{sec:ews}). 

\section{Results}

\subsection{Anomalies in different ice core proxies in response to volcanic eruptions}
\label{sec:volc_signal}

\begin{figure}
\includegraphics[width=0.99\textwidth]{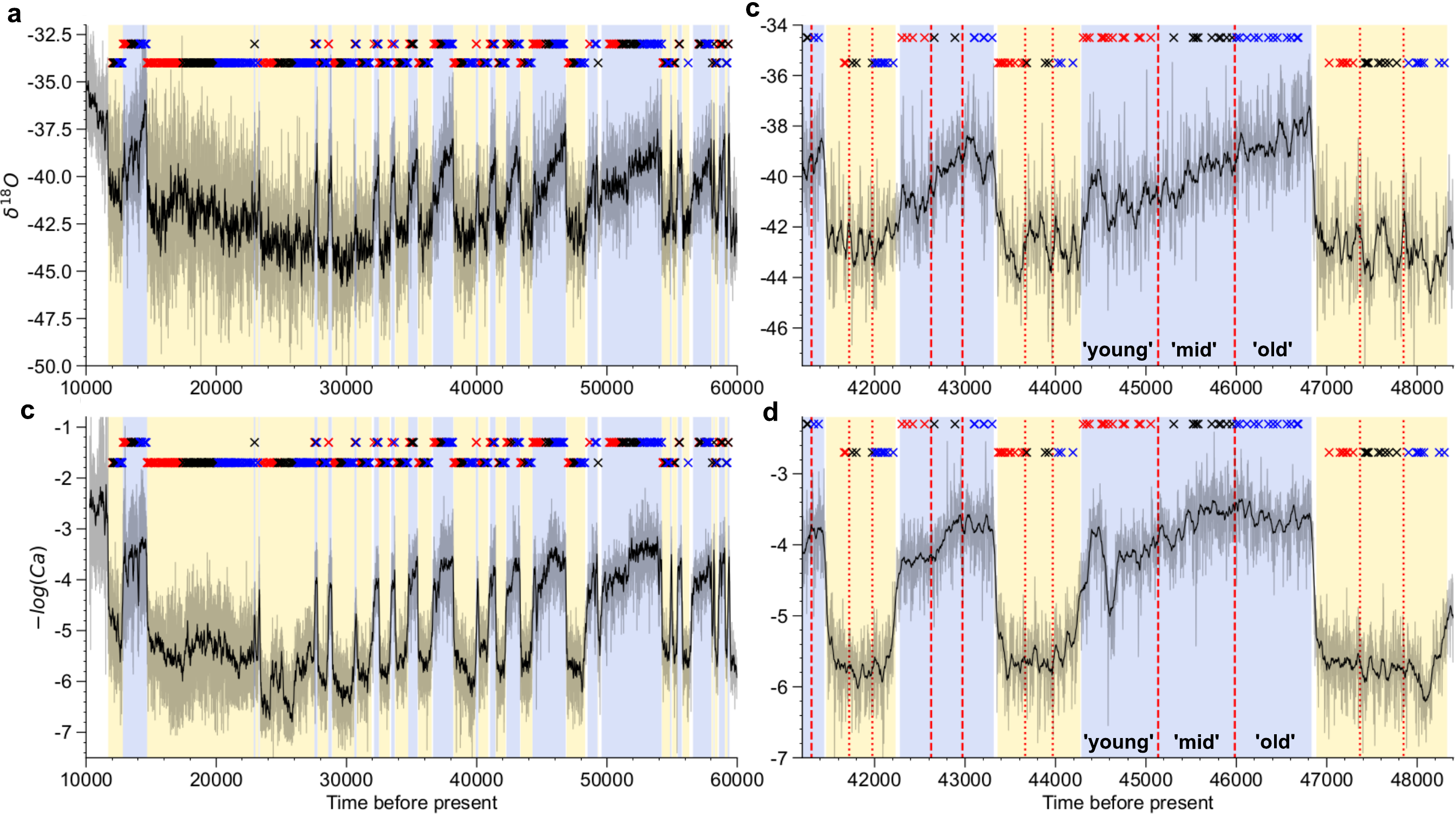}
\caption{\label{fig:timeseries} 
Two of the ice core proxy records used in this study, as well as the record of volcanic eruptions (crosses above the time series). Shown are the NGRIP $\delta^{18}$O ({\bf a}) and $Ca$ ({\bf c}) records over the time period 11.7-60 ka, as well as zoom-ins on a shorter period {\bf b,d}. A separation of the time period into warm and cold phases of DO cycles is given by shadings in grey (GI) and yellow (GS). The vertical dashed and dotted red lines in {\bf b,d}, as well as the coloring of the crosses marking volcanic eruptions, show the division of each GI and GS into three segments of equal duration. This is used here to analyze how the proxy variability and the climate response after volcanic eruptions changes as the climate progresses from the beginning (first of three segments) towards the end (third of three segments) of a GI or GS period. 
}
\end{figure}

Before turning to the analysis of CSD and EWS, the nature of the proxy records and their anomalies following volcanic eruptions is discussed. 
The time series of two proxies, as well as the timings of the volcanic eruptions are shown in Fig.~\ref{fig:timeseries}.
$\delta^{18}$O is the most commonly used record, and is a proxy for temperature. $\lambda$ (not a proxy as such) is a direct measure for relative changes in accumulation rate, when considering time scales shorter than a few millenia (see Sec.~\ref{sec:ews_significance}). The dust record measures the concentration of fine mineral particles, which are 1.5-2 microns in size on average in Greenland during the last glacial \cite{RUT03}. The signal reflects atmospheric conditions (wind speed and direction, moisture, and rainout) regulating transport and deposition efficiency, as well as conditions at the source area, i.e., more wind, drier conditions, and sparser vegetation lead to more uplift of particles. 

Unlike dust, the other proxies are concentrations of soluble impurities (ions) measured by absorption or fluorescence methods. $Ca$ is mostly the soluble part of the mineral dust, and thus highly correlated with it. 
$Na$ is mostly of sea salt origin and modulated by sea ice cover. It can indicate open ocean conditions, but during cold climates also can be uplifted from brines and frost flowers on the sea ice. 
$NH_4$ is of terrestrial origin and can indicated changes in vegetation as well as biomass burning (forest fires) \cite{LEG16}. 
$NO_3$ has been proposed as a proxy for solar activity \cite{TRA12,LAL24} (not of interest here), but also as result of various sources, including lightning, fossil fuel combustion, soil exhalation, biomass burning and ammonia oxidation \cite{SAV07}, and shows a sensitivity to temperature and accumulation \cite{ROE02}. Finally, $SO4$ is dominated by direct deposition from volcanic events, with minor terrestrial and marine biological sources. 
All impurities are also influenced by the atmospheric conditions, as the deposited concentration depends on changes in circulation and the hydrological cycle.

\begin{figure}
\includegraphics[width=0.99\textwidth]{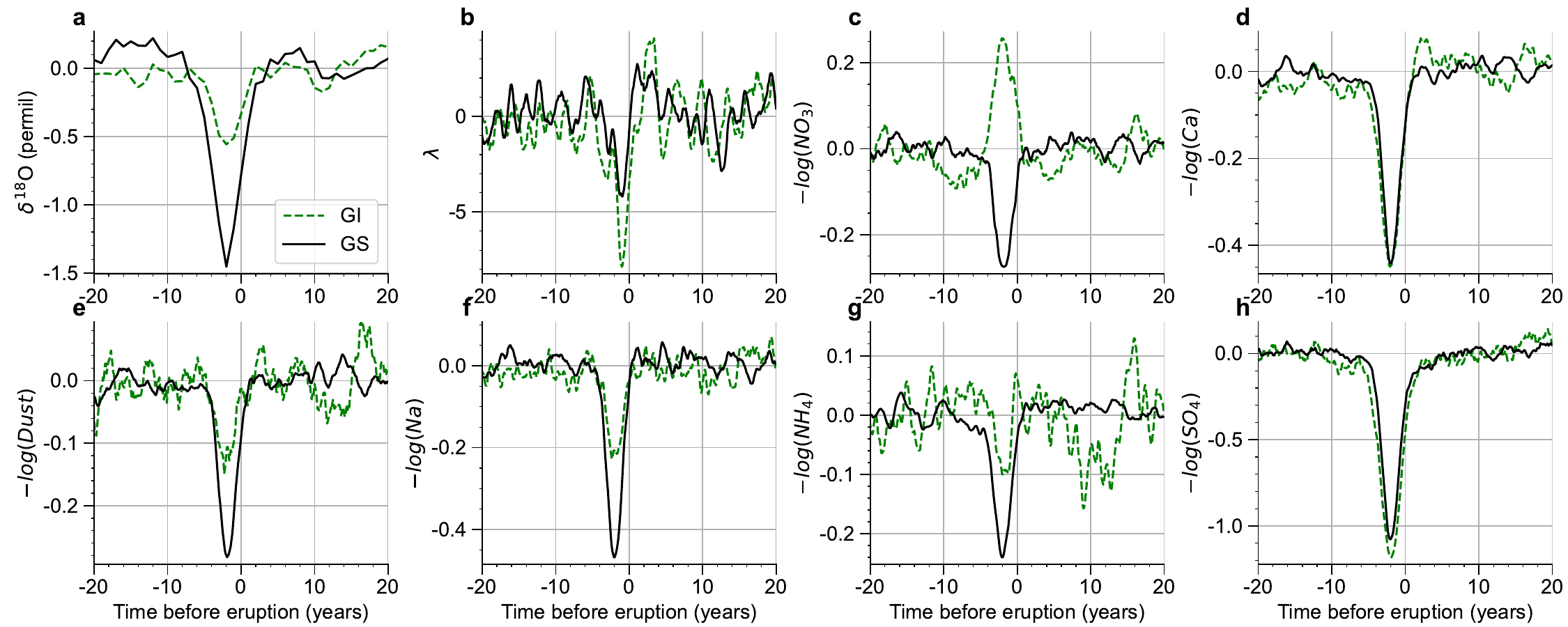}
\caption{\label{fig:gi_gs} 
Average anomalies in eight ice core proxies associated with volcanic eruptions in the interval 11.7-60 ka. The eruptions are divided according to whether they occurred during stadial (GS, black solid line) or interstadial (GI, green dashed line) periods.
For the impurity records in panels {\bf c-h}, the quantity shown is $- (\ln C - \ln C_0)$, i.e., minus the logarithm of the concentration $C$ anomalies with respect to the mean $C_0$ in the 50 years prior to an eruption. Thus, a volcanic anomaly value of, say, -0.5 corresponds to an {\it increase} in $C$ above the baseline $C_0$ by a factor of $e^{0.5} \approx 1.65$. For $\lambda$, the quantity shown is $(\frac{\lambda}{\lambda_0} - 1)\cdot 100$, i.e., the percentage change anomalies of $\lambda$ with respect to the baseline value $\lambda_0$. 
}
\end{figure}

Figure~\ref{fig:gi_gs} shows the average anomaly of the proxies in response to volcanic eruptions, where the data set is divided into eruptions occurring during the cold, stadial periods (GS) of the DO cycles and the milder, interstadial periods (GI) (see Sec.~\ref{sec:methods_eruptions}). All proxies show a clear anomaly. $\delta^{18}$O and $\lambda$ feature a negative anomaly \cite{LOH24c}, i.e., a cooling and drying consistent with the negative radiative forcing anomaly and the temperature dependence of atmospheric moisture capacity \cite{ROB94}. The $SO_4$ anomaly comes directly from the sulfuric acid aerosols emitted by the eruption, which is the forcing agent causing the climatic anomaly. 
The $SO_4$ signal increases in proportion with the magnitude of the eruption, and is included here mostly to demonstrate the constant average magnitude of eruptions (following section). 

Volcanic signals in the other impurities have been noted before, such as for $Ca$ and dust \cite{PAL91,DEA03}, but their interpretation is not straightforward. 
In general the proxies should be sensitive to post-eruptive changes in the state of the atmosphere, ocean and biosphere, in line with their usual interpretation described above. 
They should also carry a direct imprint of volcanic drying, since the post-eruptive decrease in precipitation \cite{LOH24c} leads to a larger proportion of dry deposition of aerosols (the effect of wet deposition remaining the same in terms of concentrations) and thus an overall increase in their concentration recorded in snow. Volcanic drying also likely leads to reduced rainout during transport, and (for terrestrial sources) increased uplift. This should lead to an increase, but likely not as pronounced as is observed (Fig.~\ref{fig:gi_gs}c-h) since the accumulation only decreases by about 4-7\% \cite{LOH24c}.

Importantly, in addition to a climatic signal the impurities can also carry a direct signature of the eruption itself. First, some eruptions are large and close enough to deposit volcanic ash at the ice core site, some of which can contribute to the insoluble dust signal \cite{PLU20,PLU23}. 
It may get partly dissolved and contribute to the soluble impurities signal, such as $Ca$. 
Second, the presence of volcanic ash and aerosols may lead to alterations in other impurities. Reactions with sulfuric acid can occur already in the atmosphere, e.g., with ammonia to create $NH_4$ \cite{LAN92,GFE14}, or by depleting precursors of $NO_3$ \cite{LEG90,LAJ93},
or by dissolving part of the background dust to form gypsum that contributes to the $Ca$ signal \cite{STE97b,DEA03}. 
Over long time periods, such reactions can also occur post-depositionally in the ice \cite{BAC21}, affecting the dust and impurity concentration where carbonates in dust form gypsum under the action of sulfuric acid \cite{BAC18,EIC19}. The presence of large concentrations of volcanic deposit has also been suggested to favour post-depositional migration of ions over long time periods \cite{STE97,TRA09}, and 
in particular a migration of $NO_3$ away from a volcanic layer \cite{ROE00,ROE02}. 

Here it is assumed that these non-climatic effects on the volcanic proxy anomalies are unaffected by the progression of the climate towards a loss of stability during a GS or GI. This is underpinned by the observation that the average size of the volcanic eruptions and their deposits (evidenced by the $SO_4$ spike) is unchanged during this progression (as shown further below). 
With this assumption, one can interpret changes in the average volcanic signal along this progression as an actual change in the climate response of the particular domain (atmospheric, oceanic, biosphere) that is represented by the respective proxy. This can be done without knowing the relative proportion of the climatic and non-climatic component.




Before turning to the evolution of the climate response {\it within} a GI or GS, a brief note is warranted about the difference in mean volcanic signal in GI periods versus GS overall. In \cite{LOH24c} it was shown that $\delta^{18}$O has double the anomaly in GS, and the reverse being the case for $\lambda$ (Fig.~\ref{fig:gi_gs}a,b). Such state-dependency is also observed for some impurities. 
One would expect a larger volcanic cooling in GS compared to GI (as suggested by $\delta_{18}$O) to result in more pronounced drying. This drying is on the one hand consistent with the observed larger increase in the deposition of dust, $Na$, and $NH_4$, all showing more than twice the anomaly in GS (Fig.~\ref{fig:gi_gs}e,g). 
But, as discussed in \cite{LOH24c}, the smaller reduction in $\lambda$ suggests the opposite, and thus the reasons for the observed state-dependency are more nuanced and go beyond cooling and drying. 
There is no GI-GS contrast in $SO_4$ (Fig.~\ref{fig:gi_gs}d,h), indicating that the eruptions are of similar magnitude in GI and GS (see \cite{LIN22,LOH24c}). There is also no GI-GS contrast in $Ca$ (despite the clear difference in dust), which may mean that the volcanic $Ca$ peak is dominated by effects of the actual volcanic ash/aerosol, and not by the climate response. Most interesting is $NO_3$, which shows a negative log-anomaly in GS and a positive log-anomaly (i.e. depletion) in GI (Fig.~\ref{fig:gi_gs}c). 
A depletion has been noted before for individual eruptions, and was attributed to artifacts from reactions with the volcanic deposit in atmopshere and ice \cite{LEG90,LAJ93,ROE00,ROE02}. But the opposite anomaly in GS suggests that there should be at least some non-trivial climatic component. 
To more accurately quantify the state-dependency, one needs to correct for differences in the average age of GS periods during 11.7-60 ka compared to GI periods \cite{LOH24c}, which leads to different degrees of signal preservation. Such an analysis is beyond the scope of the present paper, 
and is not necessary to analyze the climate response {\it within} individual GS and GI periods.

\subsection{Evidence for CSD in the climate response to volcanism}
\label{sec:volc_csd}

By dividing the eruptions into subsets according to how closely they occurred before a transition from GS to GI (and vice versa), it can now be determined whether the average climatic response becomes more sluggish as a transition is approached. The GS and GI periods contain a total of 526 and 220 eruptions, corresponding to a return period of 53.0 and 79.9 years, respectively. The higher frequency of recorded eruptions for GS in this data set has been noted before \cite{LOH24c}, and may be due to different mean atmospheric conditions and sulfate background noise levels during GS and GI. It is, however, not relevant for this study. 
Each GS and GI period is divided into three segments of equal duration (see Fig.~\ref{fig:timeseries}b,d). Taking together the youngest, middle and oldest segment of each GS, there are 185, 181 and 160 eruptions, respectively. For GI, the numbers are 74, 75 and 71. These numbers are consistent with a constant frequency of eruptions throughout the individual GS and GI periods.

The youngest segments in each GS and GI are closest to the purported bifurcation. Even though they vary greatly in their duration, the baseline hypothesis here is that each DO cycle essentially experiences the same progression of changes in the climate state, i.e., 
in the case of GS, starting from a state characterized by a fully collapsed AMOC and maximum NA sea ice cover, and thereafter progressing until the TP where sea ice cover is abruptly lost and the collapsed AMOC state is diminishing in stability until it resurges. 
It is thus sensible to divide each GS and GI into even segments regardless of their total length, and then lump corresponding segments together across DO cycles. 
In other words, if for example one GS lasts 3000 years and another only 300 years, the middle segments both approximately sample the same climate state on its way to destabilization, even though one lasts 1000 years and ends 1000 years prior to a transition, while the other lasts only 100 years and ends 100 years before a transition. 


\begin{figure}
\includegraphics[width=0.99\textwidth]{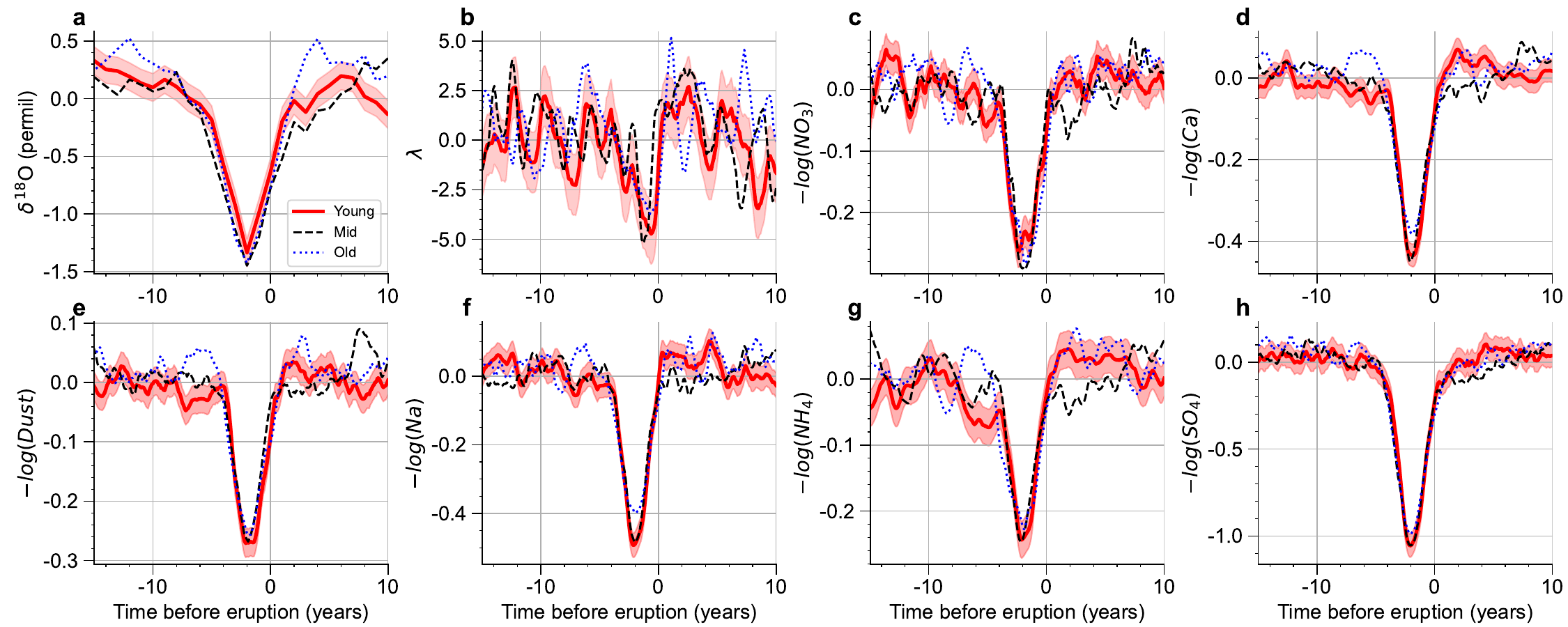}
\caption{\label{fig:volc_gs} 
Average volcanic anomalies in the eight proxies for the eruptions occurring during GS, divided into three subsets according to whether they happen at the beginning of a GS (labelled 'Old'), in the middle of a GS ('Mid') or towards the end of a GS before the abrupt transition ('Young'). For the latter category, the shading shoes the uncertainty in the average signal, given by the standard deviation of the mean signal in the 50 years prior to the eruption. 
}
\end{figure}

The average volcanic anomalies in the three segments are shown for GS in Fig.~\ref{fig:volc_gs}, and clearly the climate response captured by the proxies is the same regardless of whether an eruption occurred close to or far away from the abrupt warming transitions. Importantly, the magnitude and duration of the eruptions does not change, as the average shape and size of the sulfate spike relative to the baseline sulfur level is very similar in the three segments. 
Thus, it is a reasonable assumption that the average climate perturbation remains unchanged as the abrupt transitions at the end of individual GS/GI are approached. 

\begin{figure}
\includegraphics[width=0.99\textwidth]{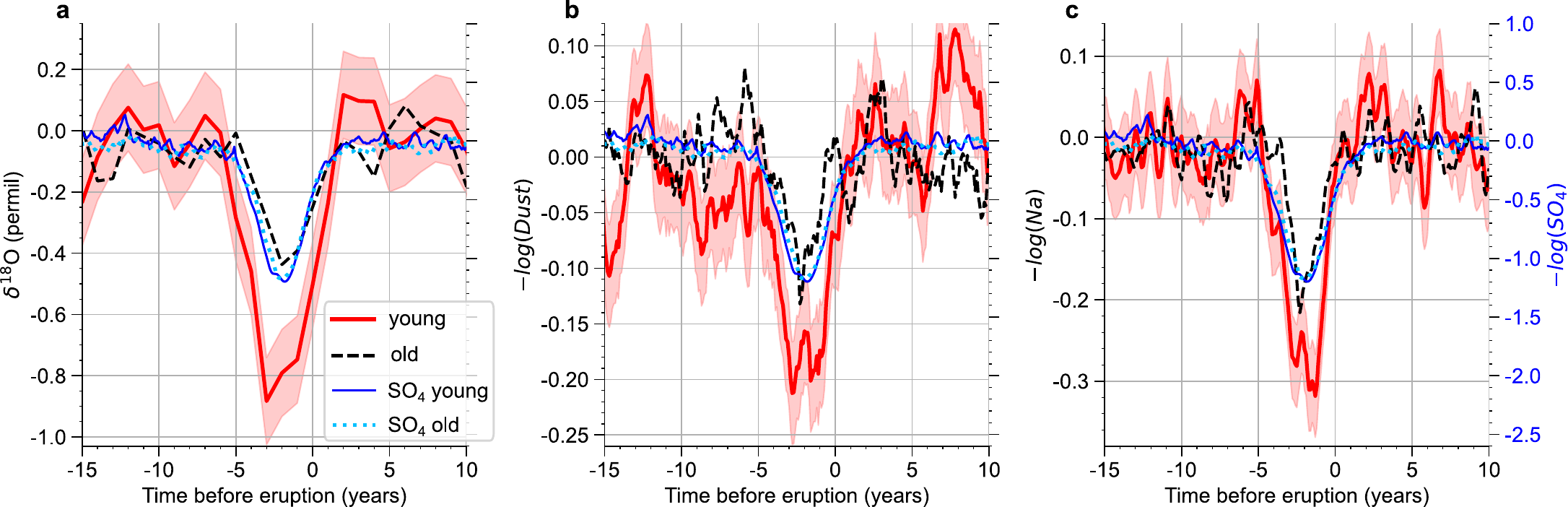}
\caption{\label{fig:volc_gi} 
Average volcanic anomalies of the proxies $\delta^{18}$O ({\bf a}), dust ({\bf b}) and $Na$ ({\bf c}) for eruptions occurring during the GI periods. The thick solid curve (red) and shading (one standard deviation of the mean signal) is the signal averaged over all eruptions that occur during the youngest third of a GI. The dashed curve is the average signal for the remaining eruptions in the older two thirds of each GI. Also shown in all panels is the average volcanic sulfate signal in the youngest (solid blue curve) and older (blue dotted line) parts of the GIs. 
}
\end{figure}

For GI, Fig.~\ref{fig:volc_gi} shows that in the three proxies $\delta^{18}$O, dust and $Na$, the anomalies associated with eruptions in the youngest segment are clearly stronger and longer lasting compared to eruptions further away from the TP. At the same time, the volcanic $SO_4$ peak stays the same (solid and dotted blue curves). 
Note that the true average volcanic climate perturbation is shorter than the $SO_4$ spike suggests, as there is uncertainty in the assignment of a definite ice core depth to the start of a volcanic eruption \cite{LIN22}. While the absolute duration of the climatic response also cannot be known precisely, the response after the younger eruptions appears to last about two years longer. 
This can be interpreted as evidence for CSD before the abrupt cooling transitions of DO cycles. 
The fact that the remaining proxies show no increase in the volcanic response (Fig.~S1) is not unexpected. The extended center manifold may be oriented in such a way that some proxies correspond to degrees of freedom that are orthogonal to it, and thus relax quickly back towards their equilibrium values before the main movement on the center manifold has occurred. It could also be that some proxies simply do not have a strong enough climatic signal in their volcanic anomaly, but mostly are related to the material of volcanic deposition itself. 


\subsection{Statistical early-warning signals by changes in proxy variability}
\label{sec:ews}

It is now tested how the direct test for CSD compares with statistical EWS, by considering the average increase in proxy variance throughout a GS or GI period. Each GS/GI is divided in 5 segments of equal length. Here more than three segments can be chosen since the data is not restricted to volcanic eruptions,  giving better statistics and a more detailed look close to the transitions where CSD may be seen. Each segment is linearly detrended, and then for a given DO cycle the variance in each segment is divided by the variance in the first (oldest) segment. The variance ratios in the five segments (the first being 1 by definition) are then averaged over the population of 22 DO cycles. Cycles with GS or GI periods shorter than 100 years are discarded to avoid large statistical fluctuations. 

This yields the average evolution of the variance over a GS or GI for each proxy, relative to the variance at the start of the GS/GI periods. To obtain an estimate for the trend over the entire GS/GI including uncertainty, statistical fluctuations are removed by fitting a straight line using least-squares, while taking into account statistical errors in the means of the five segments (obtained by bootstrap resampling with replacement). See Fig.~S2 for more detail. 

\begin{figure}
\includegraphics[width=0.75\textwidth]{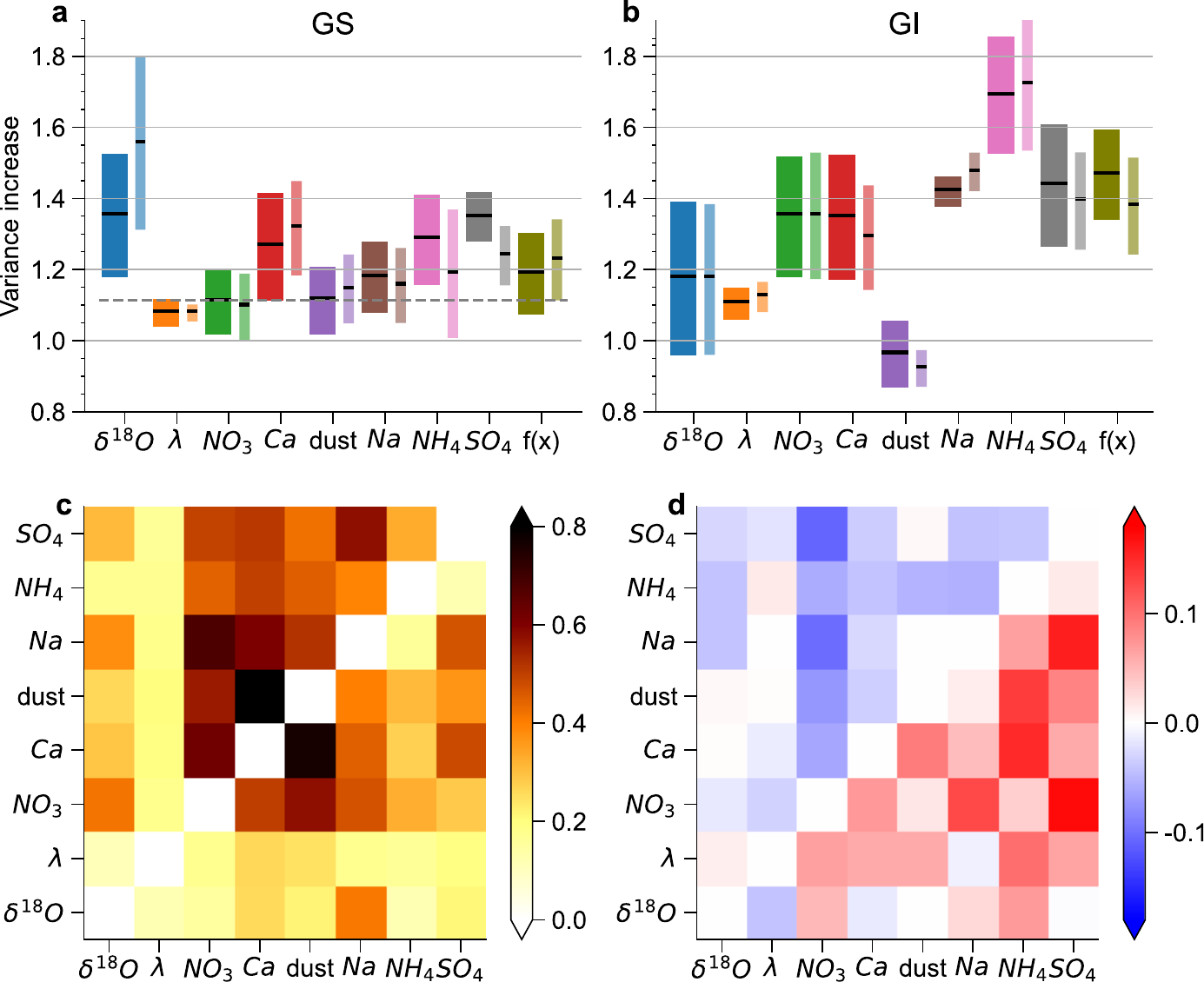}
\caption{\label{fig:variance} 
{\bf a, b} Average increase in variance of the different proxies over the GS ({\bf a}) and GI ({\bf b}) periods. The bars are 95\% confidence intervals obtained from the least-squares linear fit that is used to estimate the variance increase. The right-most bar in both panels is the variance increase of a collective variable $f(x)$ (linear combination of all proxies) obtained by the diffusion map algorithm (see main text).
The gray dashed line in ({\bf a}) is the upper bound for significance in GS (Sec.~\ref{sec:ews_significance}), taking into account proxy non-stationarity due to changing accumulation rate. 
{\bf c} Spearman correlation of the proxies in GS (upper left triangle) and GI (lower left triangle). The data used to calculate the correlation is only from the youngest third of each GS and GI. 
{\bf d} Difference in Spearman correlation of data in the youngest third minus the oldest third of each GS (upper left triangle) and GI (lower right triangle). 
}
\end{figure}

In Fig.~\ref{fig:variance} the variance increase from segment one to segment five obtained in the linear fit is shown. All proxies show a clear increase in variance in GS and GI, with the exception of dust in GI. The trend of $\delta^{18}$O in GI is quite uncertaint and not significant at 95\%. While the trends in GI are on average slightly higher, in GS they seem more consistent across the proxies. The trend of $\delta^{18}$O is largest in GS, and further increased when preprocessing with a 50-year high-pass filter. This agrees with previous analyses of EWS in $\delta^{18}$O, where a significant signal was only found in the high frequency bands \cite{RYP16,BOE18}. Otherwise, the results here do not change much when using high-pass filtered data. In GI, the strongest variance increase is for $NH_4$. 


It is unclear how to best judge the overall significance of the EWS given the various levels of variability increase in the different proxies, some of which are highly correlated. To address this, a scalar observable can be constructed from all proxies, which represents the ``critical'' climate mode associated with CSD. 
To obtain such an observable, consider that the state of a system driven by noise is given by a probability density $P(\mathbf{X}(t) = \mathbf{x} | \mathbf{X}(0) = \mathbf{x_0}) \equiv P(\mathbf{x},t | \mathbf{x_0})$ evolving in time. The evolution is governed by the Fokker-Planck (or transfer/Perron-Frobenius) operator $\mathcal{L}$, and can be written as a decomposition in eigenfunctions of $\mathcal{L}$
\begin{equation}
P(\mathbf{x},t | \mathbf{x_0}) = \sum_{n=0}^{\infty} c_n(\mathbf{x_0}) \psi_n (\mathbf{x}) e^{\lambda_n t}. 
\end{equation}
The first sub-dominant term $\psi_1 (\mathbf{x})$ with $\lambda_1<\lambda_0 = 0$ is the mode with slowest relaxation towards the equilibrium distribution $\psi_0 (\mathbf{x})$, assuming that the system resides in one metastable state before the TP, i.e., there are no noise-induced transitions. As discussed in \cite{LUC24,LOH25b}, an observable that describes the slowest degree of freedom can be obtained from the corresponding eigenfunctions of the backward Kolmogorov (or stochastic Koopman) operator $\mathcal{L}^*$, which is the adjoint of $\mathcal{L}$. 
In particular, the eigenfunction $\phi_1 (\mathbf{x})$ of $\mathcal{L}^*$ is an observable that increases along the direction of the slow relaxation mode \cite{LOH25b}. As a result, the variance of $\phi_1 (\mathbf{x})$ measured along a trajectory that experiences CSD due to an impending bifurcation is expected to increase. 

The eigenfunction is estimated here with the diffusion map algorithm \cite{COI05}. It can be viewed as a form of non-linear principle component analysis that returns the values of collective coordinates or components (which are the approximations of the eigenfunctions of $\mathcal{L}^*$) evaluated on the data points. To be able to detect the ``critical'' mode, one needs data generated relatively close to the bifurcation. Here, the data from the youngest third of each GS or GI is used (see Sec.~\ref{sec:methods_timeseries}).
After the mode $\phi_1 (\mathbf{x})$ is obtained on the set of data points, an explicit expression can be constructed for instance by the following linear approximation. 
The data points that extremize the mode are identified by choosing the top 5\% largest and 5\% smallest values of $\phi_1$. Then, averages of the original coordinates $\mathbf{x}$ in these two samples are performed. By subtracting the two averages coordinate-wise, a vector is obtained onto which new data can be projected \cite{LOH25b}. I.e., an observable is defined that is a linear combination of the variables. 

The resulting vector for GS is $[0.592, 0.360, 0.974, 1.000, 0.928, 0.971, 0.744, 0.767]^T$, and for GI it is 
$[0.553, 0.475, 0.759, 1.000, 0.982, 0.867, 0.418, 0.712]^T$. The order of the components is as in the horizontal axes of Fig.~\ref{fig:variance}a,b. The contribution of $\delta^{18}O$ and $\lambda$ is smaller compared to the impurities. This agrees with the weaker correlation of $\delta^{18}O$ or $\lambda$ and the other proxies compared to the correlation among impurities (Fig.~\ref{fig:variance}c).
In GS, the impurities contribute more equally and also $NH_4$ contributes strongly. This is in agreement with a generally stronger correlation of the impurities in GS. The evolution of the variance of these observables $f(\mathbf{x})$ over the average GS and GI is shown on the right-hand side of Fig.~\ref{fig:variance}a,b. There is a clear increase in variance, but not more pronounced than in individual proxies. It seems representative of the average variance increase over the collection of variables. 
The generally higher variance increase in GI as EWS is consistent with a more clear increase in autocorrelation across the proxies (Fig.~S3-4), as well as increasing trend in cross-correlation in GI - as would be expected in a bifurcation of coupled sub-systems \cite{DEK18,LOH21b} - versus a decreasing trend in GS (Fig.~\ref{fig:variance}d). 

\subsection{Influence of changing accumulation rate and ice core layer thickness}
\label{sec:ews_significance}

The error bars in Fig.~\ref{fig:variance}a,b represent the statistical uncertainty after averaging 22 DO cycles, where in each DO cycle there can be random trends in variance due to the correlated climate proxy variability. 
Under this uncertainty the increases appear significant for almost all proxies. But the significance needs to be further scrutinized for potential systematic effects inducing trends in proxy variability. Most notably, this is non-stationarity that arises because the effective sample spacing in ice core proxies depends on the accumulation rate at the time of snow deposition and on the subsequent thinning over time of the annual layers due to ice flow.
Because the climatic signals are smoothed by different processes across certain depth intervals in the snow, firn and ice, this changes the variability over time. The main processes are mixing of snow layers by wind-driven snow redistribution, 
physical diffusion processes in vapour in the firn, physical diffusion in the ice, and mixing in the measurement apparatus due to melting of finite amounts of material. 

The efficacy of all these smoothing processes inreases for lower accumulation rate. Processes in ice and measurement apparatus become stronger with increasing depth in the ice core due to ice flow thinning. The layer thickness record $\lambda$ captures the combined effect of accumulation rate and ice flow. The ice flow contribution can be approximately isolated by assuming that on longer time scales during the last glacial period - in particular in 60-12 ka - the mean accumulation rate did not change significantly, as evidenced by previous NGRIP accumulation reconstructions via ice flow modeling and other techniques \cite{KIN14}. Comparing the 
The mean layer thickness in GS-1 (Younger Dryas) is 2.6 times larger compared to GS-17.2 (at around 60 ka), which is hence attributed to ice flow thinning only. Comparing GI-1 (the Bølling-Allerød) to GI-17.2 (at around 59 ka), $\lambda$ changed by a factor of 2.5. Expressing the up-core increase in layer thickness as $\lambda(t+T) = \lambda(t) e^{\alpha T}$, the thinning exponent is $\alpha \approx 0.02 kyr^{-1}$ in both cases. 
During the average GS (GI), lasting 1326 (874) years, $\lambda$ thus increases due to ice flow by a factor of $f_{ice}^{GS} = 1.027$ and $f_{ice}^{GI} = 1.018$, respectively, and thus only about 2 percent. 

\begin{figure}
\includegraphics[width=0.9\textwidth]{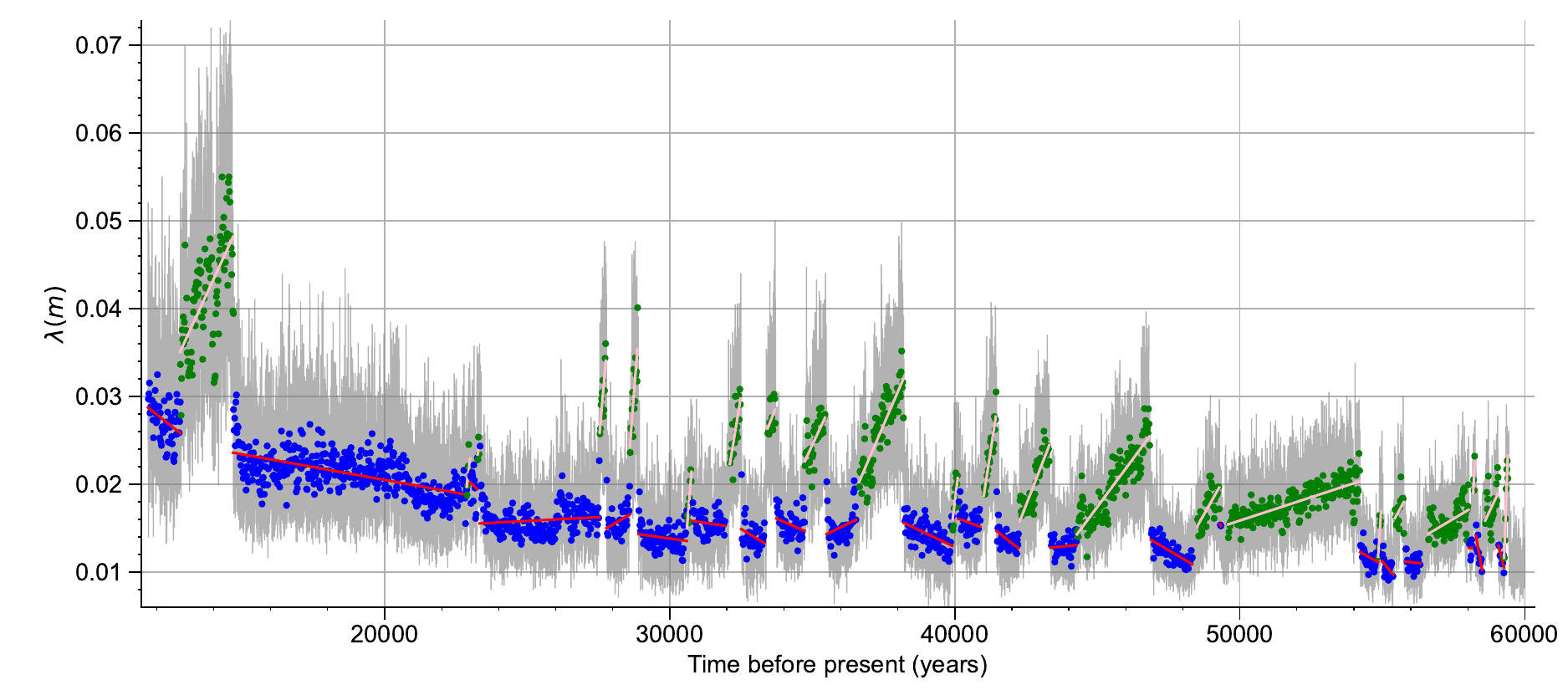}
\caption{\label{fig:lambda} 
Record of the NGRIP layer thickness in the period 12-60 ka. The high resolution record is in gray and the colored dots are 20-year averages. Also shown are least-squares linear fits in each GS and GI.
}
\end{figure}

However, linear fits to $\lambda$ within each GI and GS reveal a much larger change (Fig.~\ref{fig:lambda}). For GS, 19 out of the 22 events show an increase in $\lambda$, with a median rate of 0.15~cm per kyr. For an average layer thickness of 1.5~cm during GS, $\lambda$ thus increases by a factor of $f_{acc}^{GS} = 1.133$ within the average GS duration. 
For GI all 22 events show a decreasing $\lambda$ trend with a median 1.2 cm per kyr. At an approximate average layer thickness of 3.0cm (valid for the beginning of GIs), $\lambda$ changes by a factor of $f_{acc}^{GI} = 0.65$ during an average GI. 

\begin{figure}
\includegraphics[width=0.75\textwidth]{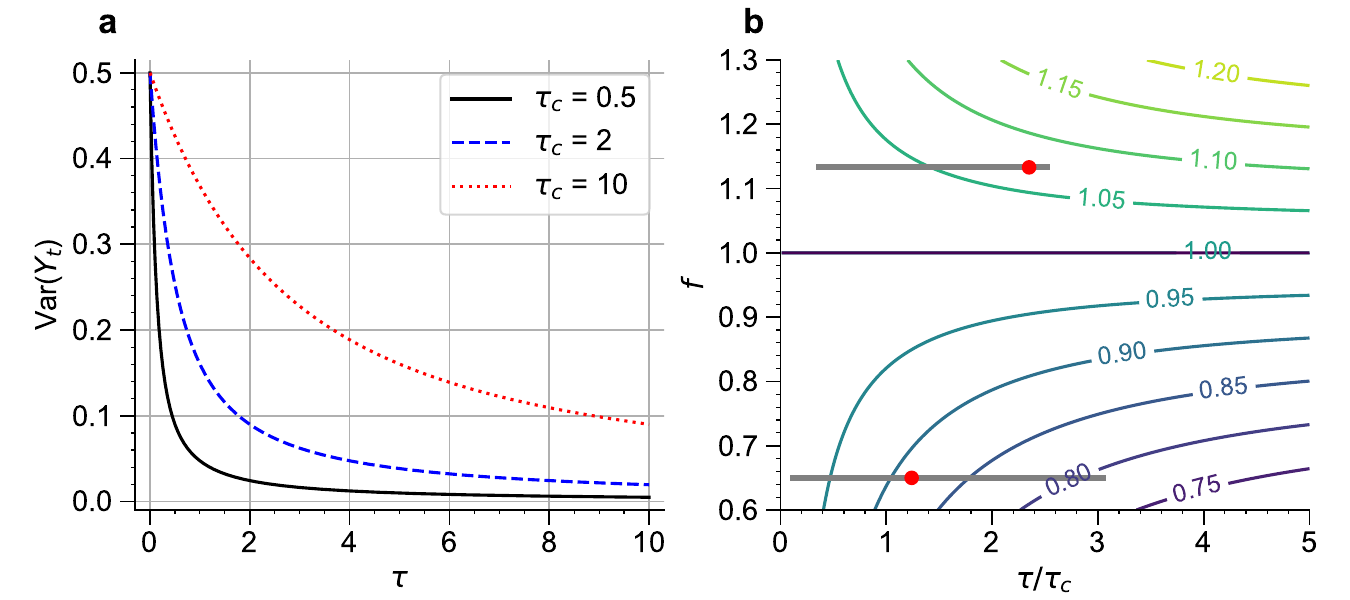}
\caption{\label{fig:csd_theory} 
{\bf a} Variance of a time-averaged Ornstein-Uhlenbeck process as function of the averaging time $\tau$, shown for different values of the correlation time $\tau_c$. Shown is the variance as in Eq.~\ref{eq:variance}, but divided by $\sigma^2 / \theta$. For $\tau \to 0$ the quantity goes to $1/2$, which then corresponds to the variance of the unaveraged process $\sigma^2 / (2\theta)$, regardless of $\tau_c$.
{\bf b} Expected change in variance (Eq.~\ref{eq:var_change}) for a time-averaged Ornstein-Uhlenbeck process when the averaging time goes from $f \tau$ to $\tau$ (contours). The gray bars (at positive $f$ for GS and negative $f$ for GI) indicate estimated ranges for $\tau / \tau_c$ for the impurity records. The red dots are the corresponding estimates for the $\delta^{18}O$ proxy. 
}
\end{figure}

Hence, in GS the variance is expected to increase because the accumulation rate tends to increase. In contrast, in GI the accumulation decreases strongly over time and thus a variance increase due to CSD is partially masked. 
It is now estimated how much the variance would seem to change during a typical GI/ GS, assuming that each proxy is subjected to an averaging process where the averaging time changes by a factor $f$. 
This averaging process is an aggregate of the different diffusion processes mentioned above. A parsimonious model for the climate records within a GS or GI, and with linear trends removed, is the Ornstein-Uhlenbeck (OU) process $X_t$ \cite{HAS76}
\begin{equation}
dX_t = -\theta X_t dt + \sigma dW_t, 
\end{equation}
with a standard Wiener process $W_t$ and a correlation time $\tau_c = \theta^{-1}$. 
The observed proxy record is then the time-averaged process 
\begin{equation}
 Y_t^\tau = \tau^{-1} \int_{t-\tau}^{t} X_{t'} dt'
\end{equation}
where the averaging time $\tau$ 
is modulated in each GS/GI by the change in accumulation rate or layer thickness. The variance is 
\begin{equation}
Var (Y_t^\tau) = \mathbb{E} \left[ \left(\tau^{-1} \int_{t-\tau}^{t} X_{t'} dt'\right)^2 \right] = \tau^{-2} \mathbb{E} \left[\int_{t-\tau}^{t} \int_{t-\tau}^{t} X_{t'} X_{s'} dt' ds' \right] = \tau^{-2} \int_{t-\tau}^{t} \int_{t-\tau}^{t} \mathbb{E} [X_{t'} X_{s'}] dt' ds' 
\end{equation}
Assuming the OU process is in statistical equilibrium, its covariance is $\mathbb{E} [X_{t} X_{s}] = \frac{\sigma^2}{2 \theta} e^{-\theta |t - s|}$ and thus 
\begin{equation}
\label{eq:variance}
Var (Y_t^\tau) = \tau^{-2} \int_{t-\tau}^{t} \int_{t-\tau}^{t} \frac{\sigma^2}{2 \theta} e^{-\theta |t' - s'|} dt' ds' = \frac{\sigma^2}{\theta^3 \tau^2} (\theta \tau + e^{-\theta \tau} -1).
\end{equation}
This function is shown for three values of $\tau_c = \theta^{-1}$ in Fig.~\ref{fig:csd_theory}a. Layer thickness and averaging time are inversely related: When $\lambda$ changes to $f \lambda$ during a GS or GI, the averaging time goes from $f \tau$ to $\tau$. Hence, the change in variance over the duration of a GI or GS period is 
\begin{equation}
\label{eq:var_change}
\frac{Var (Y_t^\tau)}{Var (Y_t^{f \tau})} = \frac{f^2 (\theta \tau + e^{-\theta \tau} - 1)}{f \theta \tau + e^{-f \theta \tau} - 1}, 
\end{equation}
which only depends on $f$ and $\theta \tau  = \tau / \tau_c$. In the limit $\tau / \tau_c \to \infty$, i.e., when the averaging time is much longer than the correlation time, 
the function converges to $f$. For $\tau / \tau_c \to 0$ the function approaches 1. These are upper and lower limits for the change in variance. The effect of layer thinning alone would be somewhere between 1 and $f\approx 1.02$, and is thus negligible. 

For a better estimate between these limits, a range of estimates for $\tau / \tau_c$ in the different proxies may be obtained. $\tau$ is derived from the effective diffusion length of a given proxy in the ice core. For $\delta^{18}O$ a diffusion length of 8~cm has been estimated at GI-1 \cite{GKI14}, i.e. in the youngest part considered here. Dividing by the mean $\lambda \approx 2.2$cm in all GI periods yields $\tau \approx 3.63$, and for GS $\tau \approx 5.39$ using the mean $\lambda \approx 1.48$cm. 
The soluble impurities are subjected to dispersion in the measurement apparatus. A lower limit (excluding dispersion during melting and debubbling) for the averaging length in NGRIP is 1.0 to 1.2 cm \cite{ERH22}. Since there are further processes, such as physical diffusion by migration in grain boundaries \cite{FAR10,NG21} and wind re-working of the snow \cite{CASA20}, an estimate of 2-3cm is used here. 
This yields ranges for the averaging of $\tau \approx 0.91 - 1.36$ in GI and $\tau \approx 1.35 - 2.02$ years in GS. 
The $\lambda$ record is not smoothed by physical processes, but by imprecision in the layer counting. This should also increase down-core, but it is hard to quantify and thus $\lambda$ will not be considered here. 

The correlation times $\tau_c$ are estimated by fitting an exponential curve to the autocorrelation function of the records during 1000-year periods in the youngest GS and GI, respectively. For $\delta^{18}O$ this yields $\tau_c = 2.30$ (GS) and $\tau_c = 2.92$ (GI). The impurities show a large range of $\tau_c$. In GS the maximum is $\tau_c = 3.62$ for $Ca$, and the minimum is $\tau_c = 0.80$ for $NO_3$. 
In GI, except for $NO_3$ (here $\tau_c = 0.44$) all correlation times are longer and reach $\tau_c = 7.27$ and $\tau_c = 9.40$ for $Ca$ and Dust, respectively. 

From this, rough estimates between the limits $\tau / \tau_c \to 0$ and $\tau / \tau_c \to \infty$ of the change in variance due to averaging can be given. Fig.~\ref{fig:csd_theory}b shows contour lines of the function in Eq.~\ref{eq:var_change} in the space spanned by $f$ and $\tau / \tau_c$. 
For the observed $f$ mainly from accumulation changes the gray bar is the range of $\tau / \tau_c$ estimated for the impurities, and the red dot is the estimate for $\delta^{18}O$.
The bar and dot for positive (negative) $f$ is for GS (GI). The expected variance increase is up to 1.08 for GS, but could be as little as 1.01 for impurities with longer correlation time, such as $Ca$. The observed variance increase is higher for all proxies, although not always within statistical uncertainty. The exception is $\lambda$, but here the extent of effective diffusion cannot be established.

For GI, the expected reduction in variance covers a larger range. The proxies with longest correlation time ($Ca$ and dust) may not experience much of an effect, but for the others (especially $NO_3$) it can be quite substantial (up to 0.78). Overall, the non-negligible expected variance reduction strengthens the significance of the observed variance increases in Fig.~\ref{fig:variance}b. 

The determined ranges of spurious variance changes are only rough estimates and the effect of averaging might still be underestimated, for instance if the correlation time was overestimated (it is here estimated only after diffusion had taken place in firn and to some extent in the ice). An additional factor that increases the effect of averaging is that the older ice at the beginning of a GS or GI had longer time to diffuse compared to the younger ice at the end of GS/GI. This impacts only ice diffusion (most relevant for $\delta^{18}O$). Since in the time period 12-60 ka the average GS and GI durations only account for 2-5\% of the total time spent for the signal to diffuse, this effect should be very small and is thus disregarded. 


\section{Discussion}
\label{sec:discussion}

The evidence for CSD leading up to the abrupt transitions of DO cycles was examined from the direct climate response after volcanic eruptions, as well as indirectly via statistical EWS in the natural climate variability. This was done for eight high-resolution proxy records in a single ice core. For GI there is a volcanic CSD signal in three of the proxies, i.e., a stronger and longer-lasting anomaly following eruptions that happen more closely before the abrupt cooling transitions from GI to GS. There is also an increase in variance as statistical EWS for all proxies except dust, which is noteworthy because dust does show a volcanic CSD signal. 
The significance of the EWS in the other proxies is strengthened, because there is a clear decrease in accumulation over the course of a GI, and one would expect an artifact in the opposite direction, i.e., a decrease in variability that would diminish EWS quite significantly. 
This also holds to some extent for the volcanic CSD, since one would expect the volcanic anomaly in the younger GI segments (with lower accumulation) to be more attenuated by the averaging/diffusion processes, but the opposite is observed. 
The atmospheric and postdepositional effects after volcanic eruptions that may introduce spurious impurity anomalies (as discussed in Sec.~\ref{sec:volc_signal})
should depend mainly on the sulfate concentrations over the background (i.e. the magnitude of the eruptions), which is approximately constant despite changing accumulation rate. 
The likelihood that the volcanic CSD is an artifact is diminished by the fact that it is seen for three different proxies ($\delta^{18}$O, dust and $Na$) with quite different post-depositional characteristics. 
The existence of CSD and EWS is furthermore corroborated by corresponding increases in autocorrelation and cross-correlation across the suite of proxies.

This combined evidence suggests that CSD is observed before the DO cooling transitions. This corroborates the observation of statistical EWS in $\delta^{18}$O and $Ca$ for some DO events \cite{MIT24}, even though as opposed to \cite{MIT24} the CSD observed here cannot be explained by the so-called rebound events that happen in some GI shortly before the abrupt DO cooling transitions. These are largely filtered out by the approaches presented here. A potential spurious contribution to the statistical EWS may be changes in precipitation intermittency \cite{CASA20}. If intermittency increases with decreasing accumulation rate during a GI, this could increase high-frequency variance due to aliasing of the seasonal cycle and introduce spurious statistical EWS. It is less clear whether this could explain the increase in autocorrelation, and, importantly, the observed volcanic CSD should not be affected. 

An unknown factor that may influence the volcanic CSD results would be systematic changes in precipitation seasonality over the course of a GI, along with a pronounced (not necessarily changing) seasonality of the volcanic climate response. As a hypothetical example, the distribution of precipitation during the early part of each GI could be relatively even throughout the year, but then, towards the end of GI, summer precipitation dominates due to a decrease in winter snowfall. If then the climatic response to the volcanic eruption is much different during summer compared to the yearly average (e.g. more pronounced cooling than in winter), it could lead to different amplitudes in early versus late GI of the annual climate anomaly recorded in a proxy. This mostly applies to proxies that record throughout the year, like $\delta^{18}$O, and not to proxies that only record seasonally, like the impurities. 
It should only affect the amplitude of the anomaly, and not its width, i.e., the duration of the climate relaxation back to steady state. In the presented analysis of CSD, increases in both amplitude and duration seem apparent, although the records after averaging are still too noisy to reliably determine the widths of the relevant anomalies (Fig.~\ref{fig:volc_gi}). While increasing relaxation time is the main hallmark of CSD, an increased amplitude after a given perturbation is also expected since the maximum departure from steady state should be larger the closer to the bifurcation due to flattening of the quasi-potential. 


For GS there is no volcanic CSD signal, and overall only a milder increase in variance. Since the accumulation rate is increasing during GS, a small artificial variance increase is expected. After quantifying this effect (Sec.~\ref{sec:ews_significance}) the increase in variance still seems significant for most proxies. But it does further diminish the strength of EWS in relation to what is seen in GI. This is corroborated by the absence of clear trends in autocorrelation, as well as decreasing trends in cross-correlation. 
The clearest evidence of EWS comes from the increase in $\delta^{18}$O variance, which is in agreement with some previous studies \cite{CIM13, RYP16, BOE18}, and which, importantly, has been show here to be also significant in the light of the changes in accumulation rate and ice flow. 
Taking together all lines of evidence, however, the overall confidence in the existence of a bifurcation precursor before DO warming transitions is low. It cannot fully be ruled out that other artifacts may increase the proxy variability over a typical GS, which would render the observed statistical EWS insignificant and in line with the (absent) CSD signal after volcanic eruptions. 

Weak or lacking evidence for CSD in GS is consistent with the observation that some of the DO warming transitions (and not the cooling transitions) may be triggered by the volcanic eruptions themselves \cite{LOH22}. In this case, the transition may occur before the system fully destabilizes and CSD would be seen. One may go further and take an absence of CSD as evidence that the DO warmings are not bifurcations, but instead purely noise-induced \cite{DIT10}. But there is evidence for underlying deterministic dynamics that allow a prediction of the DO warmings well in advance \cite{LOH19b}, inconsistent with a purely noise-induced mechanism. The predictability could arise if the DO cycles are self-sustained relaxation oscillations \cite{BRO90,KWA13,VET16,MIT17}. Such relaxation oscillations may in turn also be viewed as bifurcations in a fast sub-system, which should lead to CSD. 
Note that missing EWS may also be consistent with the idea that DO warming transitions could be due to rate-induced tipping \cite{LOH21b} where CSD is not necessarily expected. 




Future work can hopefully lower the statistical uncertainties when extended volcanic records of the older half of the glacial period become available. It would also be useful to investigate the newer EGRIP ice core when more proxy records of this core become published. Here, a decrease in $\delta^{18}$O high-frequency variability was found prior to DO warmings \cite{BRA25}, which challenges the existence of EWS and their straighforward detection. It is interesting to test how this compares with the variability in other proxies and the signal following volcanic eruptions. Finally, the findings presented here could be directly tested in Earth system models that can reproduce DO cycles. So far, these have been run without volcanic eruptions. Such simulations would give more confidence in the existence of a true CSD signal, and in particular the 
suitability of using volcanic eruptions for this purpose.
It would also give interesting insights on the proxies themselves and the state-dependency of climate variability and response that is suggested by the proxies (Sec.~\ref{sec:volc_signal}).



In conclusion, the viability of EWS to predict future climate TPs was examined by determining evidence for such precursors leading up to past abrupt climate changes (DO events) as recorded in ice core data. The main idea was to search for direct evidence of CSD the phenomenon underlying EWS) in the slowing climate response to volcanic eruptions. As opposed to statistical EWS, which only indirectly measure CSD via increasing noise-driven fluctuations, a direct CSD assessment is robust with respect to several unwanted effects, such as multiplicative noise, non-equilibrium dynamics, and non-stationarity of the noise process. 
Only the abrupt DO cooling transitions show evidence of CSD in some key ice core proxies, whereas leading up to DO warming transitions the climate response after volcanic perturbations is stationary in all proxies. In contrast, significant increases in variability (EWS) were inferred across the set of eight ice core proxies leading up to warming and cooling transitions. 
The only exception is the dust record in GI, where no EWS is seen, but instead CSD. 
The poor agreement of CSD and EWS may mean that the increases in variability of some proxies are due to specific climatic circumstances and not CSD, or are simply spurious for unknown reasons, despite the efforts here to account for an important proxy artifact, the systematic changes in accumulation rate. 
From the available data covering these past abrupt climate changes, no decisive confirmation can be made that simple EWS based on variance and autocorrelation of a priori chosen observables are reliable for future climate TPs.

\textbf{Data availability} 
The high-resolution NGRIP oxygen isotope record \cite{NGR04} is publicly available at http://iceandclimate.nbi.ku.dk/data/NGRIP\_d18O\_and\_dust\_5cm.xls. 
The NGRIP annual layer depth data set \cite{RAS23} used to construct the layer thickness record is available at https://doi.pangaea.de/10.1594/PANGAEA.943195. 
The NGRIP sulfate record is available in the supplement of \cite{LIN22} at https://cp.copernicus.org/articles/18/485/2022/. The high-resolution dust record was published in \cite{RAS23} and is available at \\
https://doi.pangaea.de/10.1594/PANGAEA.945447. 
The records of soluble impurities \cite{ERH22} are available at https://doi.pangaea.de/10.1594/PANGAEA.935837.

\textbf{Acknowledgements} 
The author would like to thank Anders Svensson and Georg Gottwald for insightful discussions related to this work. 

\bibliographystyle{apsrev4-1} 
\bibliography{refs} 

\end{document}